\documentclass[aps,prd,preprint,superscriptaddress]{revtex4-2}

\usepackage{ulem}
\usepackage{xcolor}
\usepackage[colorlinks=true]{hyperref}
\usepackage{soul}
\usepackage{booktabs}
\usepackage{mathrsfs}
\usepackage{amssymb}

\usepackage{amsmath,amssymb,color,epsfig}
\allowdisplaybreaks[4]

\newcommand{\be}{\begin{equation}}
\newcommand{\ee}{\end{equation}}
\newcommand{\bea}{\setlength\arraycolsep{2pt} \begin{eqnarray}}
\newcommand{\eea}{\end{eqnarray}}
\newcommand{\nn}{\nonumber}

\def\0{{\sst{(0)}}}
\def\1{{\sst{(1)}}}
\def\2{{\sst{(2)}}}
\def\3{{\sst{(3)}}}
\def\4{{\sst{(4)}}}
\def\5{{\sst{(5)}}}
\def\6{{\sst{(6)}}}
\def\7{{\sst{(7)}}}
\def\8{{\sst{(8)}}}
\def\sst#1{{\scriptscriptstyle #1}}

\begin{document}

\hypersetup{
    linkcolor=blue,
    citecolor=red,
    urlcolor=magenta
}


\title{Self-consistent neutron stars in a class of massive vector-tensor gravity}



\author{Zhe Luo}
\affiliation{Department of Physics, Key Laboratory of Low Dimensional Quantum Structures and Quantum Control of Ministry of Education, and Institute of Interdisciplinary Studies, Hunan Normal University, Changsha, 410081, China}
\affiliation{Hunan Research Center of the Basic Discipline for Quantum Effects and Quantum Technologies, Hunan Normal University, Changsha 410081, China}

\author{Shoulong Li}
\email[Corresponding author: ]{shoulongli@hunnu.edu.cn}
\affiliation{Department of Physics, Key Laboratory of Low Dimensional Quantum Structures and Quantum Control of Ministry of Education, and Institute of Interdisciplinary Studies, Hunan Normal University, Changsha, 410081, China}
\affiliation{Hunan Research Center of the Basic Discipline for Quantum Effects and Quantum Technologies, Hunan Normal University, Changsha 410081, China}

\author{Hongwei Yu}
\email[]{hwyu@hunnu.edu.cn}
\affiliation{Department of Physics, Key Laboratory of Low Dimensional Quantum Structures and Quantum Control of Ministry of Education, and Institute of Interdisciplinary Studies, Hunan Normal University, Changsha, 410081, China}
\affiliation{Hunan Research Center of the Basic Discipline for Quantum Effects and Quantum Technologies, Hunan Normal University, Changsha 410081, China}


\date{\today}

\begin{abstract}

Einstein-bumblebee gravity, as a class of massive non-minimally coupled vector-tensor theories, provides a useful framework for constraining Lorentz symmetry breaking through astrophysical observations, largely due to the existence of exact static and spherically symmetric black hole solutions. These solutions are typically obtained under the assumption that the vector-field potential vanishes everywhere once the vector field acquires a nonzero radial vacuum expectation value. However, imposing this assumption globally obstructs the construction of self-consistent compact-star solutions. In this work, we elucidate the origin of this inconsistency through a detailed analysis of the field equations and construct neutron-star configurations by abandoning the global vanishing-potential assumption. Crucially, we show that even without enforcing this condition everywhere, it is violated only in the strong-field interior region and is dynamically restored in the weak-field regime by asymptotic boundary conditions at spatial infinity. As a result, consistency with existing black-hole solutions and observational constraints is preserved. Our results establish massive vector-tensor gravity as a unified, natural, and self-consistent framework for compact objects, significantly extending its astrophysical viability beyond black holes and Solar System tests.

\end{abstract}


\maketitle

\section{Introduction}

Considering non-minimal couplings between the gravitational metric field and additional fields is a natural and well-motivated extension of general relativity (GR). Such theories have been extensively studied over the past several decades, as they may provide explanations for certain cosmological and astrophysical phenomena beyond the scope of GR without invoking exotic matter content, while simultaneously predicting a variety of novel gravitational effects~\cite{Will:2014kxa,Berti:2015itd,Shankaranarayanan:2022wbx,Nojiri:2017ncd,Clifton:2011jh,DeFelice:2010aj,Sotiriou:2008rp,Koyama:2015vza,Olmo:2019flu}. Depending on the nature of the additional degrees of freedom, these non-minimally coupled theories may involve scalar, vector, or tensor fields. When the extra field is a vector, the theory belongs to the class of vector-tensor theories of gravity.
 
The lowest-order and simplest vector-tensor theories with non-minimal couplings, originally studied by Hellings and Nordtvedt~\cite{Hellings:1973zz}, include the curvature-vector coupling terms $A^2 {\cal R}$ and $A^\mu A^\nu {\cal R}_{\mu\nu}$, where $A_\mu$, ${\cal R}$, and ${\cal R}_{\mu\nu}$ denote the vector field, the Ricci scalar, and the Ricci tensor, respectively. These terms can be also combined into a single coupling of the form $A^\mu A^\nu G_{\mu\nu}$, where $G_{\mu\nu}$ is the Einstein tensor, and can be further generalized to include higher-order curvature terms or derivative interactions of the vector field~\cite{Heisenberg:2014rta,Geng:2015kvs,DeFelice:2016cri,DeFelice:2016yws,Chagoya:2016aar}. In the absence of a potential term, the resulting vector-tensor theory describes a massless vector field, whereas the inclusion of a potential term, such as the simplest choice proportional to $A^2$, known as the Proca action~\cite{Proca:1936fbw}, renders the vector field massive and leads to a massive vector-tensor theory of gravity.

Recently, a particular class of massive vector-tensor theories with non-minimal couplings $A^\mu A^\nu {\cal R}_{\mu\nu}$, commonly referred to as Einstein-bumblebee gravity~\cite{Kostelecky:2003fs,Bailey:2006fd}, has attracted considerable attention. Unlike the standard Proca potential of the form $V\propto A^2$, for which the potential vanishes at $A^2=0$, these theories are characterized by a potential that vanishes when the vector field acquires a nonzero vacuum expectation value (VEV) $\langle A_\mu\rangle = b_\mu$. Specifically, the potential is chosen such that 
\be
V |_{A_\mu = b_\mu} = 0 \,, \quad \frac{dV}{d(A^2)} \Big|_{A_\mu = b_\mu} = 0  \,, \quad b_\mu b^\mu = \pm b^2 \,, \label{potentialcondition}
\ee
where $b$ is a nonzero constant. Under these assumptions, Lorentz symmetry is spontaneously broken due to the presence of a nonzero vector VEV selecting preferred spacetime directions. Although these requirements impose stringent constraints on the vector-field configuration, simple GR-like exact static and spherically symmetric black hole solutions can nevertheless be constructed within this framework by allowing a non-vanishing radial component of the vector field, 
\be
A_\mu = (0, A_r, 0, 0) \,, \label{vectoransatz}
\ee
and by imposing Eq.~(\ref{potentialcondition}) throughout the entire vacuum region~\cite{Casana:2017jkc}. Subsequent studies have further shown that Einstein-bumblebee gravity admits exact black hole solutions involving a richer set of physical parameters, including electric charge~\cite{Liu:2024axg}, magnetic charge~\cite{Li:2025tcd}, NUT charge~\cite{Chen:2025ypx}, and cosmological constant~\cite{Maluf:2020kgf}. This makes the theory a convenient and concrete framework for constraining spontaneous Lorentz symmetry breaking through astrophysical observations, as well as for investigating gravitationally induced quantum entanglement~\cite{Liu:2024wpa}.

However, when one attempts to extend the same assumptions~(\ref{potentialcondition})--(\ref{vectoransatz}) to the construction of stellar equilibrium configurations, fundamental difficulties inevitably arise. As a non-minimally coupled gravity theory, the vector field equation of motion (EOM) imposes a nontrivial additional constraint that is independent of, and not automatically guaranteed by, the metric field equations. This subtle but crucial point has been noticed from different perspectives in several recent works~\cite{Lessa:2025kln,Yu:2025odj,Ji:2024aeg}. In particular, Ref.~\cite{Lessa:2025kln} explicitly pointed out that some previously constructed compact star solutions in Einstein-bumblebee gravity~\cite{Panotopoulos:2024jtn,Neves:2024ggn}, obtained under the assumptions~(\ref{potentialcondition})--(\ref{vectoransatz}), do not satisfy the vector field EOM, and emphasized that the vector field equation must be solved explicitly when constructing compact stellar configurations. However, we find that Eqs.~(\ref{potentialcondition})--(\ref{vectoransatz}) rigidly fix the functional form of the vector field, such that the vector field EOM cannot, in general, be satisfied simultaneously in both the stellar interior and the exterior vacuum region. A more detailed demonstration of this incompatibility will be presented in subsequent sections, after specifying the field equations and the ansatz. These observations highlight that, without a fully self-consistent treatment of compact objects, the physical applicability of the theory would be confined to vacuum black hole spacetimes, thereby significantly constraining its relevance for realistic astrophysical systems.

Technically, there is no intrinsic difficulty in deriving compact star solutions from the Einstein-bumblebee gravity Lagrangian. Previous attempts failed primarily because of an unnecessarily strong and somewhat ad hoc theoretical assumption that condition~(\ref{potentialcondition}) be imposed throughout the entire vacuum region. Einstein-bumblebee gravity, as a massive non-minimally coupled vector-tensor theory, is a higher-derivative gravity theory, for which vacuum spherically symmetric solutions are not unique. In this context, black hole solutions arise only as special vacuum solutions for which condition~(\ref{potentialcondition}) can be consistently satisfied. For compact stars, however, different equations of state (EOSs) and central densities generically lead to different vacuum boundary conditions at the stellar surface. As a result, the exterior vacuum spacetime of a compact star may deviate from the equal-mass black hole geometry, and condition~(\ref{potentialcondition}) is therefore no longer satisfied. Nevertheless, such deviations are confined mainly to the strong-field region, since both black hole and compact star solutions necessarily share the same boundary conditions at spatial infinity. This implies that assumption~(\ref{potentialcondition}) need not be imposed a priori: it is naturally violated in the strong-field region while being dynamically enforced in the weak-field regime through asymptotic boundary conditions at spatial infinity. This treatment is well established in massive non-minimally coupled scalar-tensor gravity~\cite{Yazadjiev:2014cza,Liu:2024wvw,Li:2025gna,Yazadjiev:2016pcb,Staykov:2018hhc}. Consequently, consistency with previously obtained black hole solutions and existing observational constraints is preserved, and earlier Solar System tests of spontaneous Lorentz symmetry breaking based on exact black hole spacetimes remain physically meaningful.

In this work, our main goal is to construct neutron star equilibrium configurations within a vector-tensor theory framework by retaining the original Einstein-bumblebee Lagrangian and adopting the same vector-field ansatz~(\ref{vectoransatz}) used in the black hole constructions, in which the radial component of the vector field is allowed to be nonvanishing. Since assumption~(\ref{potentialcondition}) is no longer imposed---a feature specific to the conventional bumblebee model~\cite{Kostelecky:2003fs,Bailey:2006fd}---we will hereafter refer to the theory considered in this work simply as a class of massive vector-tensor gravity, rather than the bumblebee model in its strict sense.

The remainder of this manuscript is organized as follows: In Sec.~\ref{VTG}, we briefly review the massive vector-tensor theory of gravity considered in this work and present its EOMs. In Sec.~\ref{framework}, we first clarify why assumption~(\ref{potentialcondition}) cannot be imposed globally throughout spacetime. We then derive the modified Tolman-Oppenheimer-Volkoff (TOV) equations and the corresponding vacuum field equations for rotating neutron stars within the slow-rotation approximation, and present the associated boundary conditions. In Sec.~\ref{Results}, we present and analyze the numerical results for the equilibrium configurations and moments of inertia of neutron stars. Finally, Sec.~\ref{conclusion} is devoted to a summary and discussion of our results.

\section{Massive vector-tensor gravity} \label{VTG}

The total action $ S$ of the vector-tensor theory of gravity considered in this work is given by
\be
S = \frac{c^4}{16\pi G} \int d^4 x \sqrt{-g}  \left({\cal R} + \gamma {\cal R}_{\mu\nu}A^\mu A^\nu  - \frac14 F^2 - V(X ) \right) + S_\textup{m}  \,,  \label{einvec}
\ee
where $G$ and $c$ denote the gravitational constant and the speed of light, respectively. In the remainder of this work, we adopt geometric units by setting $G=c=1$. Here, $g$ denotes the determinant of the spacetime metric $g_{\mu\nu}$, ${\cal R}$ and ${\cal R}_{\mu\nu}$ are the Ricci scalar and Ricci tensor, respectively, and $A_\mu$ is the vector field. The field strength is defined as $F_{\mu\nu}=2\nabla_{[\mu}A_{\nu]}$, and $V$ represents the potential of the vector field with $X =A^2$. The dimensionless parameter $\gamma$ characterizes the non-minimal coupling between the vector field and the Ricci tensor, while $S_\textup{m}$ denotes the action of the matter sector. Varying the action with respect to the metric $g_{\mu\nu}$ and the vector field $A_\mu$, one obtains the EOMs
\bea
E_{\mu\nu} &\equiv& G_{\mu\nu} + \gamma \bigg[-\frac12 g_{\mu\nu} {\cal R}_{\rho\sigma} A^\rho A^\sigma +2 {\cal R}^\rho{}_{(\mu} A_{\nu)} A_\rho +\frac12 g_{\mu\nu} \nabla_\rho\nabla_\sigma (A^\rho A^\sigma) \nn\\
&\quad& +\frac12 \Box (A_\mu A_\nu) - \nabla_\rho\nabla_{(\mu} (A_{\nu)} A^\rho) \bigg] -\frac{ 1}{4} \left[2 F_{\mu\lambda} F_\nu{}^\lambda -\frac12 g_{\mu\nu}F_{\rho\sigma}F^{\rho\sigma} \right] \nn\\
&\quad& + \left[\frac12 g_{\mu\nu} V - \frac{\partial V}{\partial X} A_\mu A_\nu \right]   =8 \pi T_{\mu\nu}  \,, \label{eom1} \\
E_{A}^\nu &\equiv& \gamma {\cal R}^{\mu\nu}A_\mu +\frac12 \nabla_\mu F^{\mu\nu} - \frac{\partial V}{\partial X} A^\nu  = 0 \,, \label{eom2} 
\eea
where $G_{\mu\nu} \equiv  {\cal R}_{\mu\nu} - {\cal R} g_{\mu\nu}/2$ is the Einstein tensor, $T_{\mu\nu} = -\frac{2}{\sqrt{-g}} \frac{\partial S_\textup{m}}{\partial g^{\mu\nu}} $ is the energy-momentum tensor, $\nabla_\mu$ is the covariant derivative, $\Box =\nabla_\mu \nabla^\mu$ is the d'Alembert operator. Parentheses $(\mu\nu)$ and square brackets $[\mu\nu]$ indicate symmetrization and antisymmetrization over the enclosed indices, respectively.

\section{Basic equations} \label{framework}

After briefly reviewing the vector-tensor theory of gravity and its EOMs, we turn to the investigation of neutron stars within this framework. The interior and exterior spacetimes of a slowly and uniformly rotating star with angular velocity $\Omega$ can be described, to first order in  $\Omega$ of the star, by the Lense-Thirring metric ansatz~\cite{Lense:1918zz}
\be
ds^2 = - e^{\lambda(r)} dt^2 + f(r)^{-1} dr^2  +r^2 (d\theta^2 + r^2 \sin^2\theta d\varphi^2) -2\epsilon (\Omega - w(r)) r^2 \sin^2\theta dt d\varphi \,, \label{metric}
\ee
in Schwarzschild coordinates $(t, r, \theta, \varphi)$, where $\epsilon$ is a bookkeeping slow-rotation parameter, $\lambda$, $f$, and $w$ are functions of $r$, respectively. The angular velocity $\Omega$ is measured by an observer at rest at some point in the star. The function $w$ represents the angular velocity of the local inertial frame, acquired by an observer falling freely from infinity to the point in the star calculated to first order in $\Omega$, i.e., $\cal O (\epsilon)$. Their difference, $\Omega -w$, represents the angular velocity of the star relative to the local inertial frame. The slow-rotation approximation implies that the angular velocity $\Omega$ is sufficiently small, so that the changes induced by the rotation remain very slight.  We assume that the influence of the rotation on other metric functions, the vector field, and the density and pressure of stars is in second order in $\Omega$, and it is, therefore, neglected in this work.  
For the vector field $A_\mu$, we rewrite the ansatz~(\ref{vectoransatz}) in the form
\be
A_{(1)} = A_r dr = b \phi (r) dr \,, \label{vecfield} 
\ee
which yields 
\be
X = A^2 = b^2 \phi^2 f \,,
\ee
where $b$ is a nonzero constant.

Having specified the ansatz for the metric and the vector field, we now explain why the theory~(\ref{einvec}) cannot consistently admit both black hole and compact star solutions if assumption~(\ref{potentialcondition}) is imposed throughout the entire vacuum region. In vacuum, where the energy-momentum tensor $T_{\mu\nu}$ vanishes, imposing assumption~(\ref{potentialcondition}) requires $\phi = f^{-1/2}$. Substituting this relation, together with the Eqs.~(\ref{metric})--(\ref{vecfield}), into the field equations~(\ref{eom1})--(\ref{eom2}), one finds that the ${\cal O} (\epsilon^0)$ equations reduce from a higher-derivative system to an effective second-order system. This reduced system admits a unique solution for the metric functions, denoted by $f_\textup{vac}$ and $\lambda_\textup{vac}$, which corresponds to the exact black hole spacetime. Inside a compact star, however, the energy-momentum tensor $T_{\mu\nu}$ is non-vanishing, and the metric functions $\lambda$ and $f$ are inevitably modified by the presence of matter. The resulting interior solution $f_\textup{in}$ and $\lambda_\textup{in}$, therefore deviate from their vacuum counterparts. Since the vector field equation~(\ref{eom2}) is not directly sourced by the energy-momentum tensor $T_{\mu\nu}$ and admits only the vacuum solution under the constraint $\phi = f^{-1/2}$, the interior solution $f_\textup{in}$ and $\lambda_\textup{in}$ can no longer satisfy the vector field equation~(\ref{eom2}) under the same assumption. Consequently, in order for the theory~(\ref{einvec}) to consistently accommodate both black hole and compact star solutions, assumption~(\ref{potentialcondition}) must be abandoned.

These features are not unique to nonminimal coupled vector-tensor gravity~(\ref{einvec}) but are also shared by other higher-derivative gravity theories~\cite{Yazadjiev:2014cza,Liu:2024wvw,Li:2025gna,Li:2023vbo}. A representative example is Starobinsky gravity. Substituting the metric ansatz into the corresponding gravitational field equations generically leads to complicated higher-derivative equations. If one imposes the restrictive assumption $e^{\lambda}=f$ one can trivially recover the Schwarzschild metric as a black hole solution~\cite{Lu:2015cqa,Liu:2020yqa}. This assumption, however, is not justified for compact stars, whose exterior vacuum spacetime generically deviates from that of an equal-mass Schwarzschild black hole.

\subsection{Modified Tolman-Oppenheimer-Volkoff equations}

We now proceed to derive the TOV equations. Assuming that the stellar matter can be described as a perfect fluid, the corresponding energy-momentum tensor takes the form
\be
T^{\mu\nu} = (\rho + p) u^\mu u^\nu + p g^{\mu\nu} \,,
\ee
where $u^\mu$ is the four-velocity of the fluid, and is given by
\be
u^\mu = (u^t, 0, 0, \epsilon\Omega u^t) \,,  \quad \textup{with}  \quad u^\mu u_\mu = -1 \,.
\ee
Both pressure $p$ and density $\rho$ are functions of $r$, and their relationship is determined by the EOS
\be
p = P ( \rho ) \,. \label{eos}
\ee
Substituting the above Eqs.~(\ref{metric})--(\ref{eos}) into the EOMs~(\ref{eom1})--(\ref{eom2}), and solving order-by-order in $\epsilon$, the nonvanishing components of the ${\cal O}(\epsilon^{0})$ equations $E_{\mu\nu}$ and $E_{A}^{\nu}$ are given by
\begin{subequations} \label{zeroorder}
\bea
E^t{}_t &\equiv& \gamma b^2  \left[\frac{f^2 \phi ^2}{r^2}+\frac{4 f^2 \phi  \phi '}{r}+f^2 \phi'^2+f' \left(\frac{3 f \phi ^2}{r}+\frac{5}{2} f \phi  \phi '\right)+\frac{1}{4} \phi ^2 f'^2\right] \nn \\
&\quad& +\gamma b^2  f^2 \phi  \phi ''+\frac{1}{2} \gamma b^2  f \phi ^2 f''+\frac{f'}{r}+\frac{f+8 \pi  \rho  r^2-1}{r^2}+\frac{V}{2} = 0 \,, \label{ett} \\
E^r{}_r &\equiv& \gamma b^2  \left[\frac{f^2 \phi ^2 \left(-r^2 \lambda'^2+4 r \lambda'+4\right)}{4 r^2}+\frac{f^2 \phi  \phi ' \left(r \lambda'+4\right)}{2 r}\right] +\frac{V}{2} \nn \\
&\quad& \frac{-8 \pi  r^2 P(\rho )+f r \lambda'+f-1}{r^2} -\frac{1}{2} \gamma b^2  f^2 \phi ^2 \lambda''- b^2 f \phi ^2 \frac{dV}{dX} = 0 \,, \label{err} \\
E^\theta{}_\theta &=& E^\varphi{}_\varphi \equiv \gamma b^2  f^2 \phi  \phi ''+\frac{1}{2} \gamma b^2  f \phi ^2 f''+\lambda'' \left(\frac{1}{2} \gamma b^2  f^2 \phi ^2+\frac{f}{2}\right)+\frac{f' \left(r \lambda'+2\right)}{4 r} \nn \\
&\quad& +\gamma b^2  \Bigg[\frac{f^2 \phi ^2 \lambda' \left(r \lambda'+2\right)}{4 r}+\frac{f^2 \phi  \phi ' \left(r \lambda'+2\right)}{r}+f^2 \phi '^2 +\frac{1}{4} \phi ^2 f'^2 \nn \\
&\quad& +f' \left(\frac{3 f \phi ^2 \left(r \lambda'+2\right)}{4 r}+\frac{5}{2} f \phi  \phi '\right) \Bigg]+\frac{f \lambda' \left(r \lambda'+2\right)}{4 r}-8 \pi  P(\rho ) +\frac{V}{2} = 0 \,, \label{ethetatheta} \\
E^r{}_A &\equiv&  -\gamma b f^2 \phi  \lambda ''+\gamma b \left(-\frac{1}{2} f^2 \phi  \lambda'^2-\frac{f \phi  f' \left(r \lambda '+4\right)}{2 r}\right)-2 b f \phi  \frac{dV}{dX} = 0 \,, \label{eAr} 
\eea
\end{subequations}
where a prime denotes the derivative with respect to the radial coordinate $r$. When $\gamma = 0$ and $V = 0$, the vector field equation~(\ref{eAr}) is identically satisfied, and the gravitational field equations~(\ref{ett})--(\ref{ethetatheta}) reduce to those of GR. It is worth noting that, for $\gamma \ne 0$, the vector field equation imposes a nontrivial additional constraint on the system.
The nonzero components of ${\cal O} (\epsilon)$ EOM is $E^t{}_\varphi$, which is given by
\bea
E^t{}_\varphi &\equiv&  w'' - \frac{f(r \lambda' -3 r \gamma b^2 \phi^2 f' -8) - r f' + \gamma b^2 \phi f^2 (\phi( r \lambda' - 8) -4 r \phi')}{2 r f (1 + \gamma b^2  \phi^2 f)} w' \nn \\ 
&\quad& - \frac{16 \pi (\rho + P(\rho))}{ f (1 + \gamma b^2 \phi^2 f)} w = 0 \,.  \label{firstorder}
\eea
The zeroth-order modified TOV equations can be organized into a coupled system nonlinear ordinary differential equations (ODEs), which take the schematic form
\bea
\lambda^{\prime\prime} &=& F_1 (r, \lambda^{\prime}, f, \phi, \rho) \,, \nn\\
f^{\prime} &=& F_2 (r, \lambda^{\prime}, f, \phi, \rho) \,, \nn\\
\phi^{\prime} &=& F_3 (r, \lambda^{\prime}, f, \phi, \rho) \,, \nn\\
\rho^{\prime} &=& F_4 (r, \lambda^{\prime},  \rho) \,,  \label{zeroorder0}
\eea
and the first-order TOV equation then constitutes a second-order linear homogeneous ODE for $w$, which can be written as
\be
w^{\prime\prime} + F_5(r, \lambda^{\prime}, f, \phi, \rho) w^\prime  + F_6(r, \lambda^{\prime}, f, \phi, \rho) w = 0 \,,  \label{firstorder0}
\ee
where $F_i$ was used to simplify the representation of the equations and avoid displaying the specific complex expressions. 

We now aim to obtain the specific boundary conditions for Eqs.~(\ref{zeroorder})--(\ref{firstorder}). Near the center of the star, by considering power expansions for ($\lambda, f, \phi, \rho, w$),  we can express the behavior of the regular solution using a Taylor expansion, as follows:
\bea
\lim_{r \rightarrow 0} \lambda(r) &=& \sum_{i=0}^{\infty} \lambda_i r^i  \,, \nn \\
\lim_{r \rightarrow 0} f (r) &=& \sum_{i=0}^{\infty} f_i r^i \,, \nn \\ 
\lim_{r \rightarrow 0} \phi(r) &=& \sum_{i=0}^{\infty} \phi_i r^i\,, \nn \\
\lim_{r \rightarrow 0} \rho(r) &=& \sum_{i=0}^{\infty} \rho_i r^i\,, \nn \\
\lim_{r \rightarrow 0}  w(r) &=& \sum_{i=0}^{\infty} w_i r^i \,. \label{centercond}
\eea
Among the expansion coefficients, the central density $\rho_0$ and three quantities $(\lambda_0, \phi_0, w_0)$ remain as free parameters. For the functions to be solved, the remaining nontrivial leading-order coefficients, including $\lambda_2$, $f_0$, and $w_2$, are determined as
\bea
\lambda_2 &=& \frac{1+\sqrt{1+4 \gamma b^2 \phi_0^2}}{18\gamma} \frac{dV}{dX}\Big|_{X=X_0} -\frac{1}{9}V(X_0)+ \frac{16\pi}{9} (\rho_0 +2 P(\rho_0)) \,, \nn \\
f_0 &=& \frac{\sqrt{1 +4 \gamma b^2 \phi_0^2} - 1}{2\gamma b^2 \phi_0^2}  \,, \quad w_2 = \frac{8\pi}{5} w_0 (\rho_0 + P(\rho_0)) \,, \quad  X_0 = f_0 \phi_0^2 b^2 \,. \label{centercond1}
\eea
The free parameters are subsequently fixed by imposing the asymptotic boundary conditions through matching to the exterior solution at spatial infinity. For a given central density $\rho_0$ and a specific EOS, once the appropriate values for ($\lambda_0, \phi_0, w_0$) are chosen, Eqs.~(\ref{zeroorder0})--(\ref{firstorder0}) are integrated outward from the center of the star until reaching the star’s surface with radius $R$, where the pressure becomes zero, i.e., 
\be
p(R) = 0 \,. \label{surfacecond}
\ee

\subsection{Vacuum equations}

Once the integration reaches the stellar surface at $r=R$, the obtained values of $(\lambda, f, \phi, w)$ are taken as the initial conditions for the exterior integration. These are fixed by the continuity conditions at the stellar surface,
\bea
\lambda_\textup{in}(R) &=& \lambda_\textup{ext}(R) \,, \nn \\
f_\textup{in}(R) &=& f_\textup{ext}(R) \,,  \nn \\
\phi_\textup{in}(R) &=& \phi_\textup{ext}(R) \,,  \nn \\
w_\textup{in}(R) &=& w_\textup{ext}(R) \,.  \label{continue}
\eea
where the subscripts ``in'' and ``ext'' denote the interior and exterior solutions, respectively. The first derivatives of these functions are likewise continuous at $r=R$, although not written explicitly here. Since the curvature invariants can be expressed in terms of the metric functions and their derivatives, this continuity guarantees that the spacetime curvature remains regular across the stellar surface.  It is worth noting that neutron-star spacetimes differ qualitatively from black-hole geometries in this respect. In certain black-hole solutions of this theory, the vector field can exhibit a coordinate singularity at the event horizon~\cite{Li:2025tcd}. By contrast, in the neutron-star configurations considered here, the vector field remains regular throughout the entire spacetime, both in the interior and in the exterior regions. 

We then continue the integration outward through the vacuum region up to spatial infinity. The vacuum field equations are given by Eqs.~(\ref{zeroorder})--(\ref{firstorder}) with $p= 0$ and $\rho=0$. The zeroth-order vacuum equations can be organized into a coupled system nonlinear ODEs, which take the schematic form
\bea
\lambda^{\prime\prime} &=& \hat{F}_1 (r, \lambda^{\prime}, f, \phi) \,, \nn\\
f^{\prime} &=& \hat{F}_2 (r, \lambda^{\prime}, f, \phi) \,, \nn\\
\phi^{\prime} &=& \hat{F}_3 (r, \lambda^{\prime}, f, \phi) \,,   \label{zeroorder0vac}
\eea
and the first-order vacuum equation can be expressed as
\be
w^{\prime\prime} + \hat{F}_4(r, \lambda^{\prime}, f, \phi) w^\prime  = 0 \,,  \label{firstorder0vac}
\ee
where $\hat{F}_i$ was used to simplify the representation of the equations and avoid displaying the specific complex expressions.  At asymptotic infinity, we can also express the behavior of the functions $\lambda, f, \phi$, and  $w$ using a Taylor expansion, as follows:
\bea
\lim_{r \rightarrow \infty}  \lambda(r) &=&  \sum_{i=0}^{\infty} \hat{\lambda}_i r^{-i}   \,,  \nn \\
\lim_{r \rightarrow \infty}  f(r) &=& \sum_{i=0}^{\infty} \hat{f}_i r^{-i} \,, \nn \\
\lim_{r \rightarrow \infty}  \phi(r) &=& \sum_{i=0}^{\infty} \hat{\phi}_i r^{-i}  \,, \nn \\
\lim_{r \rightarrow \infty}  w(r) &=& \sum_{i=0}^{\infty} \hat{w}_i r^{-i} \,. \label{inftycond0}
\eea
Substituting these expansions into the ${\cal O}(\epsilon^{0})$ vacuum equations, i.e. Eq.~(\ref{zeroorder}) with $p=0$ and $\rho=0$, and solving them order by order in ${\cal O}(r^{-1})$, one finds that the leading-order field equations take the form
\bea
E^t{}_t &\equiv& \frac12 V(\hat{X}_0) + {\cal O} (r^{-1}) \,,\nn \\
E^r{}_r &\equiv& -b^2 \hat{\phi}_0^2 \hat{f}_0 \frac{dV}{dX}\Big|_{X=\hat{X}_0} + \frac12 V(\hat{X}_0)  + {\cal O} (r^{-1}) \,,\nn \\
E^\theta{}_\theta &\equiv& \frac12 V(\hat{X}_0) + {\cal O} (r^{-1}) \,,\nn \\
E^r{}_A &\equiv& -2 b^2 \hat{\phi}_0^2 \hat{f}_0 \frac{dV}{dX}\Big|_{X=\hat{X}_0} + {\cal O} (r^{-1})\,,
\eea
where $\hat{X}_0 = b^2  \hat{\phi}_0^2 \hat{f}_0$. Requiring the leading-order vacuum field equations to be satisfied immediately implies
\be
V(\hat{X}_0) = 0 \,, \quad \frac{dV}{dX}\Big|_{X=\hat{X}_0} = 0 \,. \label{potentialcond1}
\ee
At this stage, the explicit functional form of the potential $V(X)$ has not yet been specified. We therefore assume a class of potentials for which the above conditions are satisfied at $\hat{X}_0 = b^2$, such as
\be
V(X) = \alpha \gamma^2 (X - b^2)^2 \,, \label{potentialcond2}
\ee
where $\alpha$ is a constant characterizing the mass scale of the vector field. Once this potential is specified, together with Eq.~(\ref{potentialcond1}), it follows that in the asymptotic region $r\to \infty$, the functions $\lambda, f$, and $\phi$ admit the following exact asymptotic forms:
\bea
\lim_{r \rightarrow \infty}  \lambda(r) &=& \log \left(1 - \frac{2 M}{r \sqrt{1 + \gamma b^2}} \right) \,, \nn \\ 
\lim_{r \rightarrow \infty}  f(r)  &=& \frac{1}{1 + \gamma b^2} \left(1 - \frac{2 M}{r \sqrt{1 + \gamma b^2}} \right) \,, \nn \\ 
\lim_{r \rightarrow \infty}  \phi(r)  &=&  (1 + \gamma b^2)^{\frac12} \left(1 - \frac{2 M}{r \sqrt{1 + \gamma b^2}} \right)^{-\frac12} \,, \label{inftycond1}
\eea
where $M$ represents the Noether mass~\cite{Li:2025tcd,An:2024fzf,Chen:2025ypx}. Here, we introduce the so-called Lorentz-violation parameter 
\be
\ell = \gamma b^2 \,, 
\ee
which is dimensionless and allows for a direct comparison with existing constraints from black hole and Solar System tests. Upon expressing the field equations and the near-center boundary conditions in terms of $\ell$, the constant $b$ can be completely absorbed and therefore does not appear explicitly in the equations to be solved. Substituting these asymptotic expressions further into the ${\cal O}(\epsilon)$ vacuum equations, i.e. Eq.~(\ref{firstorder}) with $p=0$ and $\rho=0$, one can determine the exact solution of the function $w(r)$ at spatial infinity, which is given by
\be
\lim_{r \rightarrow \infty}  w(r) = \Omega - \frac{2 J}{r^3} \,,\label{inftycond2}
\ee
where $J$ represents the angular momentum of the star.
The moment of inertia $I$ of the star is defined as
\be
I = \frac{J}{\Omega} \,. \label{moi}
\ee 
At the end of this section, we adopt  the symbol ``$\star$'' as a subscript to represent the dimension of various physical quantities, 
\be
M_\odot  \sim r_\star \sim   p_\star^{-\frac12} \sim \rho_\star^{-\frac12} \sim \alpha_\star^{-\frac12} \sim I_\star^{\frac13} \,, \label{dimension}
\ee
with the Solar mass $M_\odot$ serving as the reference unit. These quantities allow us to transform the Eqs.~(\ref{zeroorder})--(\ref{firstorder}) to a dimensionless form for numerical treatment. The characteristic dimensions of relevant physical quantities, normalized by the solar mass $M_{\odot}$ and expressed in centimeter-gram-second (cgs) units, are given by
\bea
r_\star &=& \frac{G M_{\odot}}{c^2} = 1.47 \times 10^{5} \textup{cm} \,, \nn\\
\rho_\star &=& \frac{c^6}{G^3 M_{\odot}^2 } = 6.18\times 10^{17} \textup{g} \cdot \textup{cm}^{-3} \,, \nn\\
p_\star &=& \frac{c^8}{G^3 M_{\odot}^2 } =5.55\times 10^{38}  \textup{g} \cdot \textup{cm}^{-1} \cdot \textup{s}^{-2} \,, \nn\\
I_\star &=& \frac{G^2 M_{\odot}^3}{c^4} = 4.34\times 10^{43}  \textup{g} \cdot \textup{cm}^{2}  \,.
\eea

\section{Equilibrium configurations and moments of inertia}  \label{Results}

Having established all the equations to be solved, namely Eqs.~(\ref{zeroorder0})--(\ref{firstorder0}), (\ref{zeroorder0vac})--(\ref{firstorder0vac}), along with the boundary conditions at the center~(\ref{centercond})--(\ref{centercond1}), at the surface~(\ref{surfacecond}), and at asymptotic infinity~(\ref{inftycond1}) and (\ref{inftycond2}), and continuity condition at the surface~(\ref{continue}), we now specify the EOS~(\ref{eos}) for the neutron stars. In this work, we adopt the realistic SLy (Skyrme Lyon) EOS~\cite{Douchin:2001sv,Haensel:2004nu} for illustration. To solve the set of nonlinear ODEs, we treat it as a boundary value problem and apply the shooting method. Note that since $\lambda(r)$ appears in the field equations only through its first derivative, the function $\lambda$ itself does not need to be solved explicitly. We match all functions to be solved, $(\lambda^\prime, f, \phi, w, w^\prime)$, at a sufficiently large radius, such that the computed gravitational mass is accurate to at least two decimal places. In the following sections, we will analyze the results for both the stellar structures under hydrostatic equilibrium and the moments of inertia of stars in the context of first-order slow rotation.

\begin{figure}[]
\includegraphics[width=0.95\linewidth]{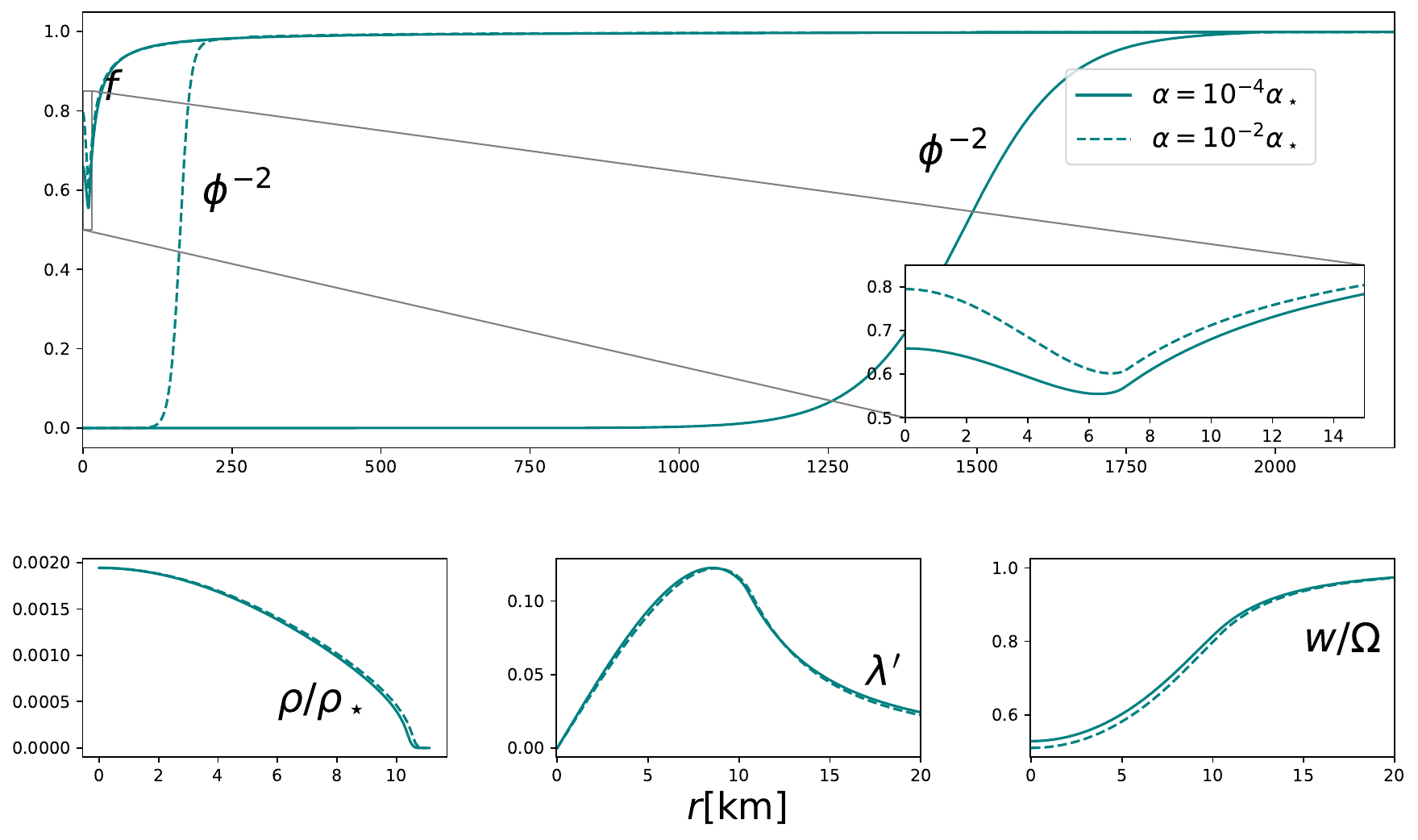}
\caption{\label{fig:0thsolution}The  plots illustrate the numerical solutions of the functions ($ f, \phi^{-2}, \rho, \lambda', w/\Omega$) in massive vector-tensor gravity with the parameters $ \alpha = 10^{-4}  \alpha_\star$ (solid lines) and $\alpha = 10^{-2} \alpha_\star $ (dashed lines). The parameter $\ell$ is fixed to $\ell=10^{-10}$. The central density is set to $\rho_0 = 1.2 \times 10^{15} \textup{g/cm}^3$, and the SLy EOS  is adopted.}
\end{figure}

To verify the consistency of the neutron star solutions obtained in this work, we first plot the metric functions ($ f, \phi^{-2}, \rho, \lambda', w/\Omega$), as functions of the radial coordinate $r$ in Fig.~\ref{fig:0thsolution}, for representative values of the parameters $\ell$ and $\alpha$. 
Motivated by the bounds $(10^{-12}, 10^{-9})$ on the parameter $\ell$ inferred from three classical Solar System tests~\cite{Casana:2017jkc}, we choose $\ell=10^{-10}$ for our numerical analysis. We consider two  values of the parameter $\alpha$ namely $ \alpha = 10^{-4}  \alpha_\star$ and $\alpha = 10^{-2} \alpha_\star $. The central density is fixed at $\rho_0 = 1.2 \times 10^{15} \textup{g/cm}^3$. For both choices of $\alpha$, the resulting neutron star mass is $1.51 M_{\odot}$, while the corresponding radii are $7.48 r_\star$ (11.0 km)  and $7.58 r_\star$  (11.1 km), and the moments of inertia are $34.92 I_\star$ and $34.25 I_\star$. From the upper panel of Fig.~\ref{fig:0thsolution}, one can clearly see that $\phi^{2} f$ deviates from unity both inside the neutron star and in the exterior strong-field region, while it approaches unity only in the weak-field regime and asymptotically at spatial infinity, as required by the boundary condition~(\ref{potentialcond1}). Moreover, the deviation becomes more pronounced as the potential parameter $\alpha$ decreases.

\begin{figure}[]
\includegraphics[width=0.95\linewidth]{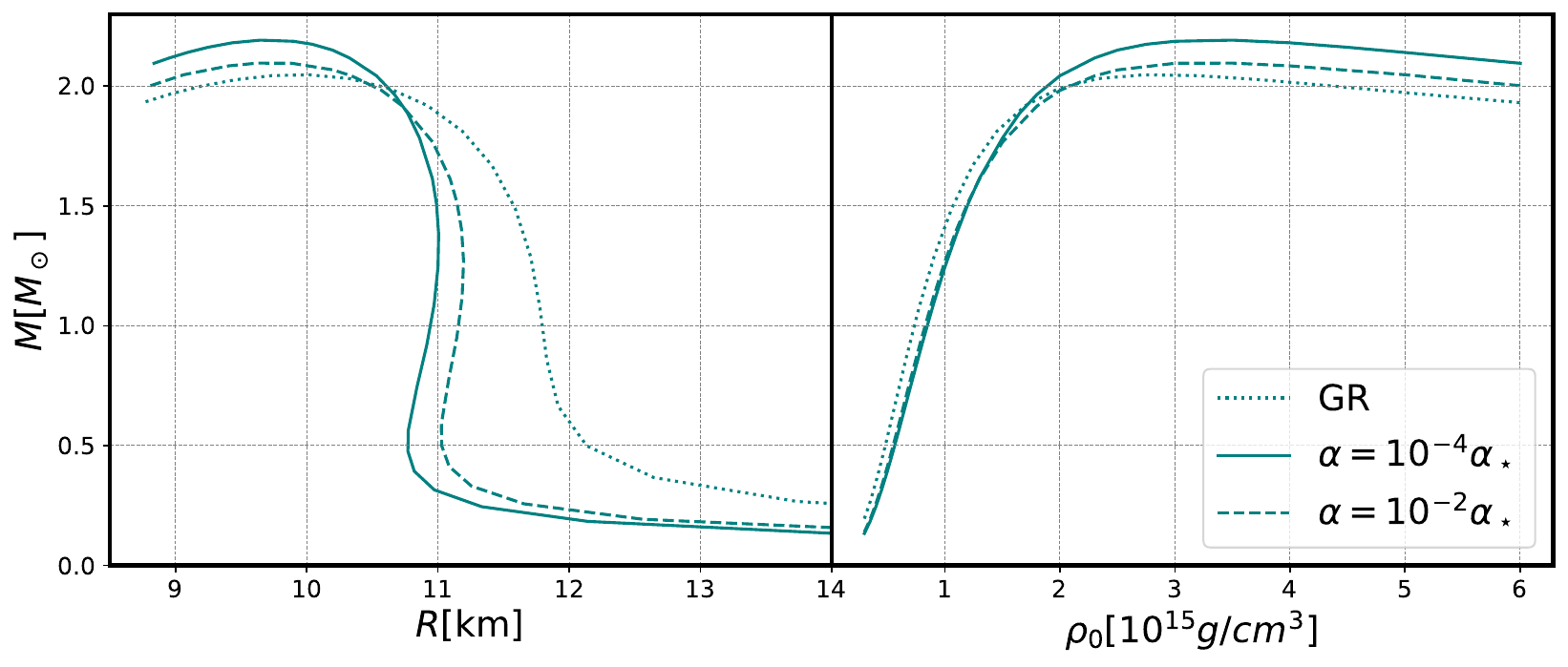}
\caption{\label{fig:MRrelation}Mass-radius ($M$-$R$, left) and mass-central density ($M$-$\rho_0$, right) relations for neutron stars described by the SLy EOS in massive vector-tensor gravity. The parameter $\ell$ is fixed to $\ell = 10^{-10}$. Different line styles correspond to $\alpha = 10^{-4}\alpha_\star$ (solid), $\alpha = 10^{-2}\alpha_\star$ (dashed), and GR (dotted). }
\end{figure}

Next, we plot the mass-radius $M-R$ and mass-central density $M-\rho_0$ relations for the SLy EOS over a range of central densities $\rho_0$, considering two representative values of the parameter, $\alpha = 10^{-4}\alpha_\star$ and $\alpha = 10^{-2}\alpha_\star$, as shown in Fig.~\ref{fig:MRrelation}. 
This allows us to provide a more comprehensive assessment of the effects of the non-minimal coupling term and the potential on stellar configurations with different central densities. It is evident that even when the parameter $\ell$ is as small as $\mathcal{O}(10^{-10})$, the mass and radius of neutron stars can still undergo significant modifications. Compared with the GR case, neutron stars exhibit smaller masses and radii at low central densities, while larger masses and radii are obtained in the high-central-density regime. Overall, larger values of the vector-field mass parameter $\alpha$ drive the stellar macroscopic properties closer to the GR limit, whereas smaller values lead to more significant deviations from GR, approaching the behavior of the corresponding massless theory. This deviation from GR becomes increasingly pronounced as $\alpha$ decreases. Interestingly, the impact of the deviation depends on the central density: compared with GR, low-central-density neutron stars tend to exhibit smaller masses and radii, while high-central-density configurations can attain larger masses and radii. This qualitative trend is similar to that found in non-minimally coupled massive scalar-tensor gravity~\cite{Yazadjiev:2014cza,Liu:2024wvw,Li:2025gna,Yazadjiev:2016pcb,Staykov:2018hhc,Staykov:2014mwa}.

\begin{figure}[]
\includegraphics[width=0.75\linewidth]{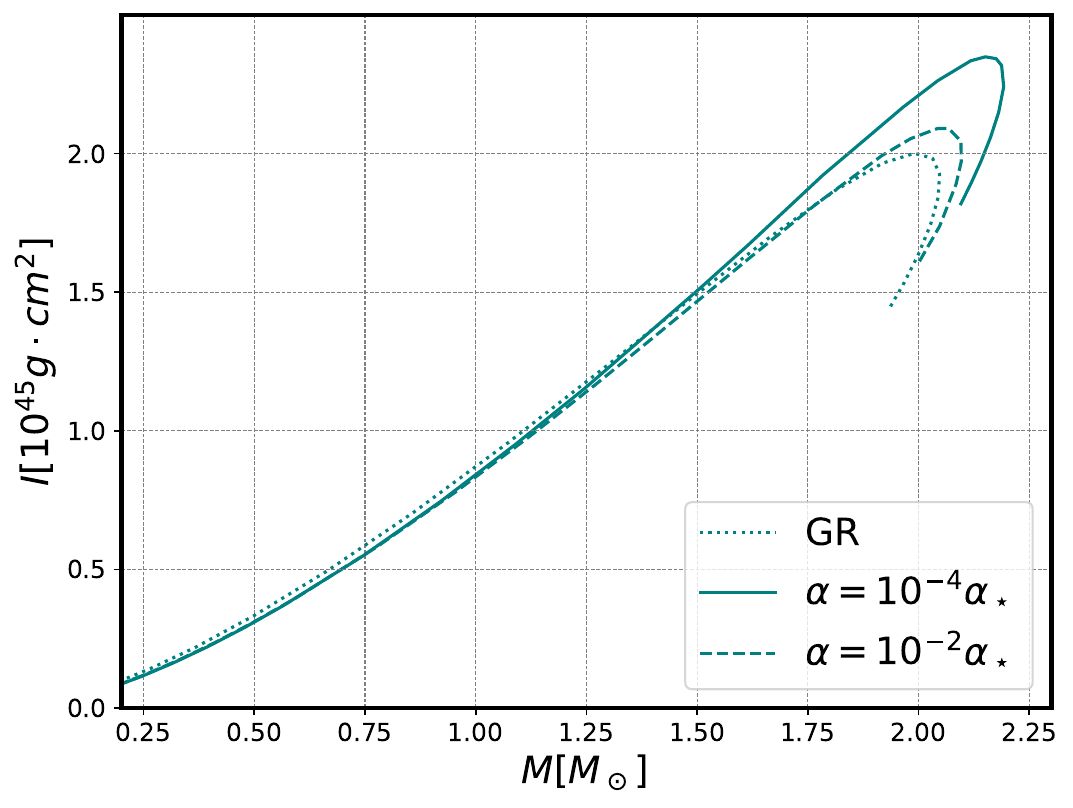}
\caption{\label{fig:IMrelation}Moment of inertia-mass ($I$–$M$) relations for neutron stars described by the SLy EOS in massive vector-tensor gravity. The parameter $\ell$ is fixed to $\ell = 10^{-10}$. Different line styles correspond to $\alpha = 10^{-4}\alpha_\star$ (solid), $\alpha = 10^{-2}\alpha_\star$ (dashed), and GR (dotted).}
\end{figure}

Finally, we discuss the effects of the non-minimal coupling term and the potential on the rotational properties of neutron stars, which can be characterized by the moment of inertia $I$ in the slow-rotation approximation, as given by Eq.~(\ref{moi}). Figure~\ref{fig:IMrelation} presents the $I-M$ relation for the neutron stars obtained previously. 
Compared with the GR case, neutron stars in massive vector-tensor gravity exhibit smaller moments of inertia at low masses and larger moments of inertia at high masses. The influence of the non-minimal coupling term is relatively weak for low-mass stars but becomes more pronounced for high-mass stars. Moreover, this effect is stronger for smaller values of the parameter $\alpha$, indicating that the non-minimal coupling has a significant impact in the strong-field region, and that the magnitude of the vector-field mass plays an important role in this regime. This behavior is also similar to that observed in non-minimally coupled massive scalar-tensor gravity~\cite{Yazadjiev:2014cza,Liu:2024wvw,Li:2025gna,Yazadjiev:2016pcb,Staykov:2018hhc,Staykov:2014mwa}.

\section{Conclusion} \label{conclusion}

In this work, we have investigated neutron stars within a class of massive vector-tensor theories of gravity, characterized by a non-vanishing VEV of the vector field. Owing to the fact that the vector field selects a preferred spacetime direction, Lorentz symmetry is spontaneously broken in this theory.  Exact static and spherically symmetric black-hole solutions had previously been constructed in this theory, for which condition~(\ref{potentialcondition}) can be satisfied globally throughout spacetime. These solutions further enabled stringent Solar System constraints on the Lorentz-violating parameters of the theory. Motivated by this  success, similar assumptions were subsequently adopted in attempts to construct compact star solutions. In Sec.~\ref{framework}, however, we  demonstrated that imposing such assumptions a priori inevitably leads to inconsistencies. We further showed that these assumptions are not fundamental requirements of the theory: instead they are  dynamically enforced in the weak-field regime through the asymptotic boundary conditions at spatial infinity. As a result, the theory remains fully compatible with existing black hole solutions and Solar System tests, while providing a unified, natural, and self-consistent framework for constructing neutron star solutions.  Notably, this theoretical observation is not limited to the construction of neutron star solutions in non-minimally coupled vector-tensor models of Lorentz-violating gravity, but can be extended to theories in which the metric is non-minimally coupled to additional antisymmetric two-form fields~\cite{Lessa:2025kln}.

Based on our calculations of the neutron star mass, radius, and moment of inertia, we arrive at two main conclusions. First, even when the Lorentz-symmetry-breaking parameter $\ell$ is tightly constrained to very small values by Solar System tests, the presence of the vector-field potential term can still induce appreciable deviations in both the mass-radius relation and the moment of inertia of neutron stars relative to their GR counterparts. Second, although the explicit functional form of the potential does not affect black hole solutions, it has a significant impact on the properties of neutron stars. More specifically, larger values of the vector-field mass parameter $\alpha$ drive the stellar macroscopic properties closer to the GR limit, whereas smaller values lead to greater deviations from GR, approaching the behavior of the corresponding massless theory. These features closely parallel those observed in massive scalar-tensor gravity in the Einstein frame, both with and without self-interaction potentials~\cite{Yazadjiev:2014cza,Liu:2024wvw,Li:2025gna,Yazadjiev:2016pcb,Staykov:2018hhc,Staykov:2014mwa}. A detailed comparison with neutron star solutions in Einstein-Kalb-Ramond gravity and in massive Brans-Dicke theory formulated in the Jordan frame with self-interaction potentials would therefore be worthwhile.

Finally, it is worth emphasizing that, although we have established the existence of neutron star solutions within a class of massive vector-tensor theories satisfying the asymptotic condition~(\ref{potentialcond1}), such models---owing to the presence of self-interacting vector fields---may still exhibit dynamical pathologies~\cite{Coates:2022qia,Barausse:2022rvg,Coates:2022nif,Coates:2023dmz,Unluturk:2023qgk,Coates:2023swo,Unluturk:2024ltf,Rubio:2024ryv}. In particular, they can be prone to instabilities associated with loss of hyperbolicity and tachyonic modes. Several approaches have been proposed to mitigate the problem of dynamical evolution~\cite{Rubio:2024ryv}, and a more complete ultraviolet completion of the theory is expected to ultimately cure or regulate these difficulties. Nevertheless, a systematic understanding of how such dynamical instabilities can be consistently resolved remains a central open challenge in this class of gravity theories.

Additional important avenues for future investigation include the following. First, it is natural to investigate whether non-minimal gravitational couplings can have a significant impact on stellar stability~\cite{Kokkotas:1999bd}. A systematic analysis of the radial and non-radial oscillation stabilities and quasi-normal modes of neutron stars within this framework would be highly valuable.  Second, since the non-minimal coupling term and the potential term can affect the macroscopic properties of neutron stars, it is natural to consider additional observables, such as the quadrupole moment and tidal Love numbers. These quantities allow one to make use of approximately EOS-insensitive universal relations, including the I-Love-Q relations~\cite{Yagi:2013bca,Yagi:2013awa}, thereby providing complementary constraints on the Lorentz-violating parameter as well as on the form and parameters of the vector-field potential. Third, a promising direction is to investigate the impact of couplings between the self-interacting vector field and the Maxwell field on neutron-star structure. Recent studies indicate that nonlinear electrodynamics can nontrivially influence magnetar shape~\cite{Suvorov:2025ffh}, suggesting that such interactions may likewise lead to novel phenomenological signatures.



\begin{acknowledgments}
We are grateful to the referee for the constructive and insightful comments on our paper.
We are grateful to Peixiang Ji, H. L\"u and Lijing Shao for useful discussions. 
S.L. and H.Y. were supported in part by the National Natural Science Foundation of China (No. 12105098, No. 12481540179, No. 12075084, No. 11690034, No. 11947216, and No. 12005059) and the Natural Science Foundation of Hunan Province (No. 2022JJ40264), and the innovative research group of Hunan Province under Grant No. 2024JJ1006, and by the Excellent Young Scholars Program of the Hunan Provincial Department of Education under Grant No. 25B0092. 

\end{acknowledgments}


\end{document}